%% file: main.tex
\documentclass[%
reprint,
superscriptaddress,
amsmath,amssymb,
aps,
prl,
]{revtex4-2}
\usepackage{graphicx}
\usepackage{dcolumn}
\usepackage{bm}
\usepackage{braket}
\usepackage{amsthm}
\usepackage{siunitx}
\usepackage{xcolor}
\usepackage{soul}
\usepackage{float}
\usepackage{xr}
\usepackage{xr-hyper}
\usepackage[colorlinks=true,linkcolor=blue,citecolor=blue,filecolor=blue,urlcolor=blue]{hyperref}

\begin{document}
	
	\preprint{APS/123-QED}
	
	\title{Angular-resolved nonlinear optical response as a probe of Lorentz violation in noncentrosymmetric materials}
	
	\author{Guilherme J. Inacio}
	\affiliation{Departamento de Física, Universidade Federal do Espírito Santo, 29075-910 Vitória-ES, Brazil}
	\affiliation{Departamento de F\'isica de la Materia Condensada, Universidad Aut\'{o}noma de Madrid, E-28049 Madrid, Spain}
	
	\author{Nathanael N. Batista}
	\affiliation{Departamento de Física, Universidade Federal do Espírito Santo, 29075-910 Vitória-ES, Brazil}
	
	\author{Wesley Spalenza}
	\affiliation{Instituto Federal do Esp\'{i}rito Santo, IFES - Cariacica, ES, Brasil}
	\affiliation{Centro Brasileiro de Pesquisas Físicas, CBPF - RJ, Brasil}
	
	\author{Humberto Belich}
	\affiliation{Departamento de Física, Universidade Federal do Espírito Santo, 29075-910 Vitória-ES, Brazil}
	
	\author{Juan Jos\'e Palacios}
	\affiliation{Departamento de F\'isica de la Materia Condensada, Universidad Aut\'{o}noma de Madrid, E-28049 Madrid, Spain}
	\affiliation{Condensed Matter Physics Center (IFIMAC), Universidad Aut\'{o}noma de Madrid, E-28049 Madrid, Spain}
	\affiliation{Instituto Nicol\'as Cabrera (INC), Universidad Aut\'{o}noma de Madrid, E-28049 Madrid, Spain}
	
	\author{Wendel S. Paz}
	\email{wendel.paz@ufes.br, corresponding author}
	\affiliation{Departamento de Física, Universidade Federal do Espírito Santo, 29075-910 Vitória-ES, Brazil}
	\affiliation{Departamento de F\'isica de la Materia Condensada, Universidad Aut\'{o}noma de Madrid, E-28049 Madrid, Spain}

	\date{\today}
	
	\begin{abstract}
		We propose a methodology to detect weak Lorentz-violating (LV) backgrounds through the nonlinear shift photocurrent in noncentrosymmetric crystals. Using a spinful Rice--Mele model, we show that a LV background induces a momentum-odd correction to the Bloch Hamiltonian that reshapes the phase of the interband dipole matrix elements. As a result, the shift conductivity develops a robust $\pi$-periodic modulation as a function of the angle of a perpendicularly applied static electric field, in contrast to a weakly $2\pi$-periodic response of the Lorentz-symmetric case. This change in angular periodicity provides a signature of LV effects which can be directly identified through a photocurrent measurement. For realistic optical intensities, the predicted signal lies in the picoampere range, which can be enhanced in a matrix of weakly interacting chains, allowing sensitivity to LV coupling strengths of the order of $\xi\sim10^{-24}\,\mathrm{C\,m}$. These results establish nonlinear optical transport as a viable probe of emergent LV effects in solid-state systems.
	\end{abstract}

	\maketitle
	
	
	Lorentz symmetry plays a central role in modern physics ensuring that the laws of nature are invariant under changes of orientation and uniform motion. Within the Standard-Model Extension (SME), this symmetry can be relaxed in a controlled way by allowing matter and gauge fields to couple to fixed background tensors, leading to small deviations from exact Lorentz and CPT invariance \cite{Colladay1997,Colladay1998,Kostelecky2004,Kostelecky2011}. In this framework, nonminimal SME-based couplings can generate Rashba-type interactions and modified Landau levels in nonrelativistic Hamiltonians \cite{bakke2014rashba,bakke2015influence,bakke2015landau}. A broad range of precision experiments has used the SME to place tight bounds on possible deviations, including tests based on cavity resonators, atomic spectroscopy, spin-precession measurements, and astrophysical observations \cite{Gabrielse1999,FermiLAT2009,IceCube2018,ALPHA2017,Muong-2_2008,Battat2007}. 
	Condensed matter systems offer a complementary perspective, as they provide controllable platforms where symmetry-breaking effects can be engineered and probed experimentally. In Weyl and Dirac semimetals, for example, tilted dispersions and parity-odd responses can mimic Lorentz-violating (LV) terms \cite{Armitage2018,Soluyanov2015,Grushin2012,Burkov2014,Xiong2015}. Yet, the influence of actual LV backgrounds on measurable solid-state observables remains largely unexplored \cite{kostelecky2022lorentz}.
	
	Nonlinear photocurrents, and especially the shift current, are known to be highly sensitive to the phase structure and Berry geometry of Bloch states in crystals that lack inversion symmetry \cite{sipe2000second,young2012first,Morimoto2016, garcia_blazquez_shift_2023, esteve_paredes_excitons_2025,pietralonga2026probing}. These responses have been measured with high precision in a variety of low-dimensional semiconductors, including transition-metal dichalcogenide nanoribbons, nanotubes, ferroelectric layers, and flexo-photovoltaic devices \cite{xue_ws2_2024,zhang_enhanced_2019,chen_bulk_2025,pal_bulk_2021,li_enhanced_2021,yang2018flexo}. Their sensitivity and experimental maturity make them ideal tools for detecting subtle symmetry-breaking perturbations that modify band geometry.
	
	\begin{figure}[b]
		\centering
		\includegraphics[width=\linewidth]{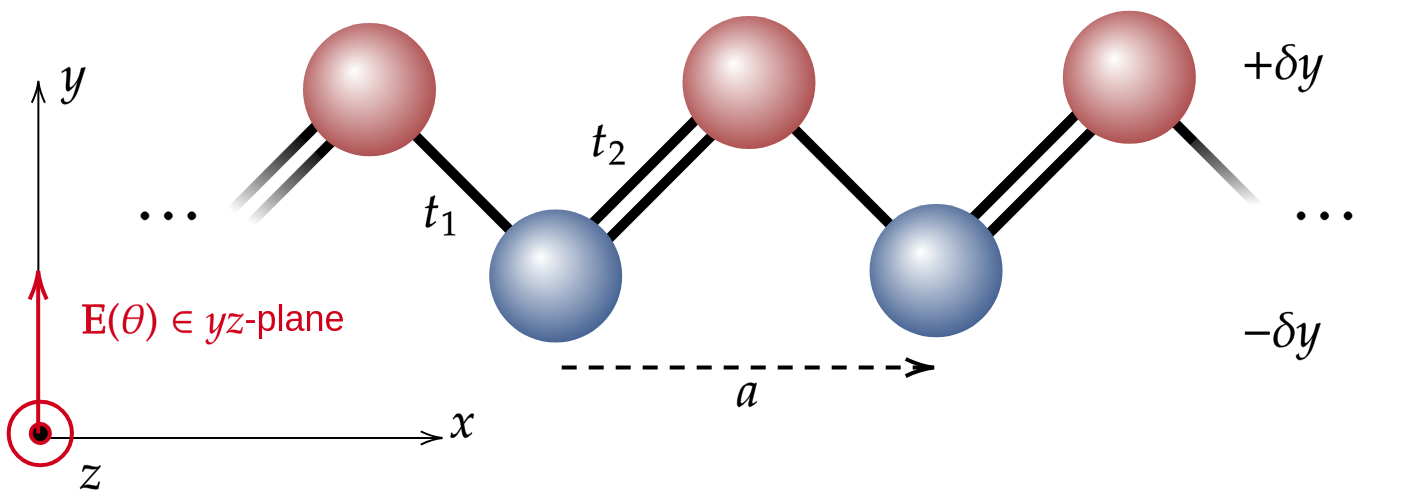}
		\vspace{-10px}
		\caption{
			Schematic of the atomic chain polar lattice described by the Rice--Mele model, with alternating hoppings $t_1$ and $t_2$ and staggered sublattice positions $\pm \delta_y$. The chain extends along $\hat{x}$, while a static electric field $\mathbf{E}_{\mathrm{dc}}(\theta)$ is applied in the transverse $yz$ plane. Rotating $\theta$ enables an angular scan of the nonlinear response, where the emergence of a $\pi$-periodic modulation in $\sigma(\theta)$ serves as a signature of the Lorentz-violating perturbation.
		}
		\label{fig:device}
	\end{figure}
	
	In this work, we propose an angular-resolved method to probe LV backgrounds through the shift photocurrent of a polar atomic chain described by a spinful Rice--Mele model and realized experimentally as an aligned bundle of weakly interacting chains. A static electric field is rotated in the transverse ($yz$) plane while the photocurrent is measured along the chain direction ($\hat{x}$), as illustrated in Fig.~\ref{fig:device}. In the Lorentz-symmetric case, the response exhibits only a weak $2\pi$-periodic angular dependence. In contrast, the LV perturbation induces a robust $\pi$-periodic modulation in $\sigma^{(2)}(\theta)$ for generic orientations of the background vector $\mathbf v$. This change in angular periodicity provides a direct experimental signature, which may be identified through Fourier analysis of the measured $I(\theta)$ or $\sigma^{(2)}(\theta)$ over several rotations of the field.
	
	\textit{Theory ---} LV effects can persist in the non-relativistic limit, as shown by Kosteleck\'y and Lane \cite{Kostelecky1999}. To capture the essential mechanism by which an LV background modifies the band geometry and the nonlinear optical response, we consider a one-dimensional noncentrosymmetric lattice described by a spinful Rice--Mele model \cite{rice1982elementary}.
	We start from a Dirac equation supplemented by a non-minimal coupling to a fixed background four-vector $v^\mu$ \cite{belich2005non,BELICH1,BELICH2,BELICH3,Jackiw}. After a Foldy--Wouthuysen reduction, the non-relativistic Hamiltonian contains orbital and spin contributions proportional to the background field \cite{bakke2014rashba,bakke2015influence},
	\begin{equation}
		H_{\mathrm{LV}}
		=
		\frac{g}{2m_e}\,
		\boldsymbol{\sigma}\!\cdot\!
		\left[
		\boldsymbol{\nabla}\times
		\left(
		\mathbf v\times\frac{\mathbf E}{c}
		\right)
		\right]
		-
		\frac{g}{m_e}\,
		\mathbf p\!\cdot\!
		(\mathbf v\times\mathbf E),
	\end{equation}
	where the first term acts in the spin channel and the second in the orbital one. These terms provide the continuum structure from which the lattice perturbation is constructed.
	
	The unperturbed crystal is modeled by the spinful Rice--Mele Hamiltonian,
	\begin{equation}
		\label{eq:HRM_spinful_new}
		H_{\mathrm{RM}}(k)
		=
		\Big[
		d_x(k)\tau_x+d_y(k)\tau_y+d_z\tau_z
		\Big]\otimes\sigma_0,
	\end{equation}
	with
	\begin{align}
		&d_x(k)=t_1+t_2\cos(ak),\\
		&d_y(k)=-t_2\sin(ak),\\
		&d_z=\Delta,
	\end{align}
	where $a$ is the lattice constant, $t_1$ and $t_2$ are the intra- and intercell hopping amplitudes, $\Delta$ is the staggered on-site potential, $\tau_i$ act on sublattice space, and $\sigma_i$ act on real spin.
	We consider a chain oriented along $\hat x$, with Bloch momentum $k$. The LV background vector $\mathbf v$ is allowed to be fully three-dimensional, while the external static electric field is restricted to the transverse $yz$ plane,
	\begin{align}
		\mathbf v &= v_0
		\left(
		\sin\beta\cos\phi\,\hat x
		+\sin\beta\sin\phi\,\hat y
		+\cos\beta\,\hat z
		\right),\\
		\mathbf E &= E_0
		\left(
		\sin\theta\,\hat y+\cos\theta\,\hat z
		\right).
	\end{align}
	With this choice, no static field is applied along the periodic direction, and the LVP is controlled by the projection of $\mathbf v\times\mathbf E$ along the chain axis.
	
	Projecting the continuum correction onto the two-sublattice Rice--Mele basis yields a $k$-odd contribution in the intercell hopping channel,
	\begin{equation} \label{eq:H_LV_1D}
		H_{\mathrm{LV}}(k) = 
		\tau_y\otimes 
		\Big[
		\lambda_{\mathrm{orb}}(k)\sigma_0 
		+ \lambda_{\mathrm{spin}}(k) 
		\left( \sin\theta\,\sigma_y+\cos\theta\,\sigma_z \right)
		\Big],
	\end{equation} 
	where 
	\begin{align}
		\lambda_{\mathrm{spin}}(k) &= 
		\frac{\hbar \xi E_0}{2m_e c}\, v_0\sin\beta\cos\phi\, k,\\ 
		\lambda_{\mathrm{orb}}(k) &= 
		-\frac{\hbar \xi E_0}{m_e c}\, v_0 
		\left( 
		\sin\beta\sin\phi\cos\theta-\cos\beta\sin\theta 
		\right)k, 
	\end{align}
	and $\xi=g/c$. The orbital term is spin independent, whereas the spin term inherits the transverse field orientation through the combination $(\sin\theta\,\sigma_y+\cos\theta\,\sigma_z)$. In both cases, the perturbation is odd in $k$ and enters the Bloch Hamiltonian through the $\tau_y$ channel, thereby modifying the phase structure of the Bloch spinors. To account for the pure static field $\mathbf{E}$ contribution, the external field is coupled to the real space positions of the two sublattices, generating a staggered onsite potential between sites $A$ and $B$ according to their distance $\pm\delta y$. This contribution is a $k$-independent term of the form $H_{\mathrm{ext}}\propto \tau_z$, which defines the baseline response in the absence of the LVP. The explicit derivation of the LVP and the static field term is given in Supplementary Material S1.  
	
	The full effective Hamiltonian is therefore
	\begin{equation}
		H_{\mathrm{eff}}(k)=H_{\mathrm{RM}}(k)+H_{\mathrm{LV}}(k) + H_{\mathrm{ext}}.
	\end{equation}
	
	\begin{figure}[t]
		\centering
		\includegraphics[width=\linewidth]{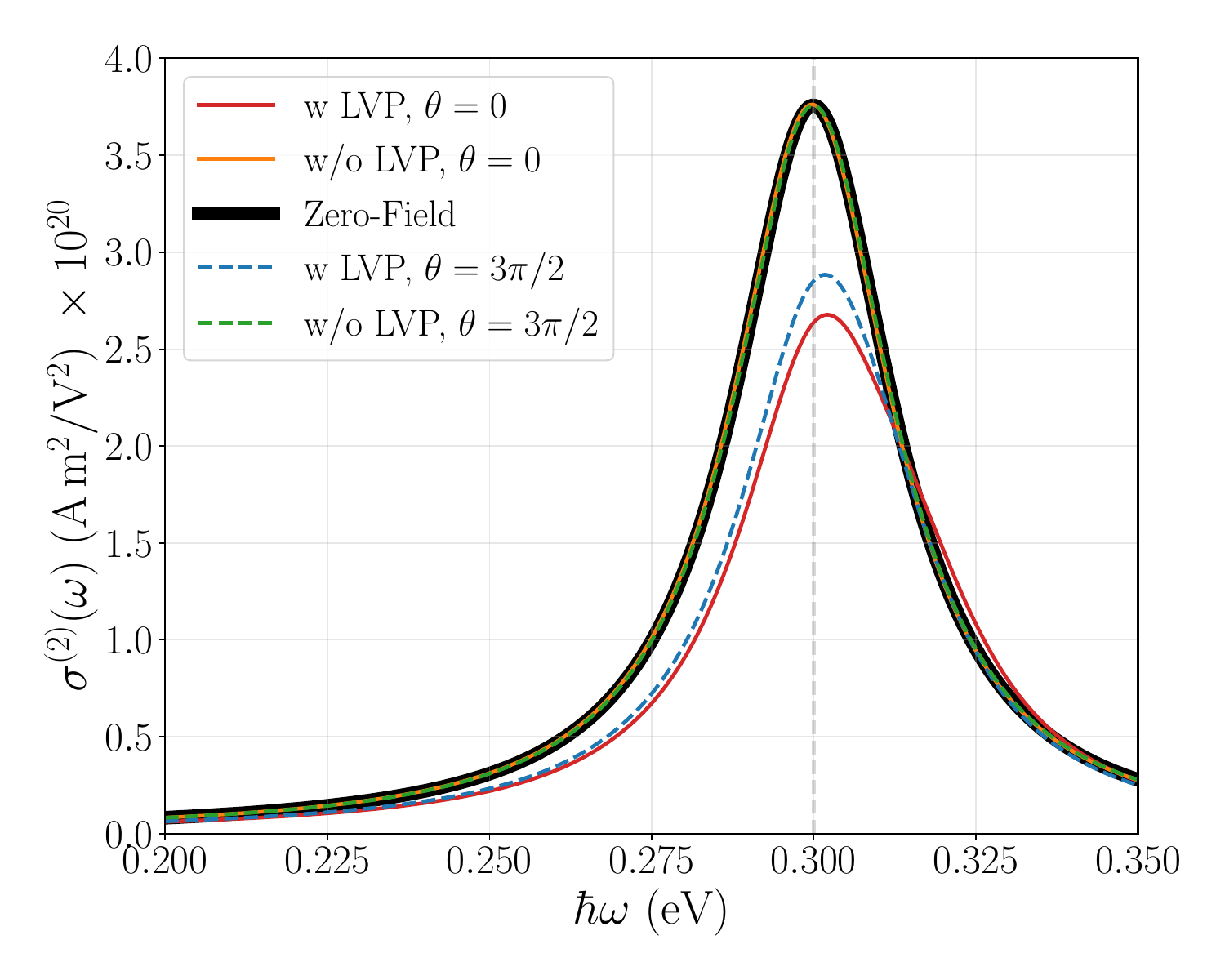}
		\vspace{-25px}
		\caption{
			Shift conductivity $\sigma^{(2)}(\omega)$ for two field orientations separated by $\pi/2$. The solid black curve corresponds to the zero-field case. In the absence of the LVP, the spectra for different angles nearly coincide, indicating a weak angular dependence. When the LVP is included, deviations appear near the band-edge resonance, revealing a pronounced dependence on the field orientation.
		}
		\label{fig:linear}
		\vspace{-10px}
	\end{figure}

	The eigenstates of $H_{\mathrm{eff}}(k)$ determine the interband dipoles and shift vectors entering the length-gauge expression for the second-order shift conductivity \cite{sipe2000second,young2012first,Morimoto2016,esteve_paredes_excitons_2025},
	\begin{align}
		&\sigma^{(2)}_{abb}(\omega,\theta)= \\
		&-\frac{\pi e^{3}}{\hbar^{2}}
		\sum_{m,n}
		\int\frac{dk}{2\pi}
		f_{mn}(k)
		R^{a}_{mn}(k)
		|r^{b}_{mn}(k)|^{2}
		\delta\big(E_{mn}-\omega\big).\notag
	\end{align}
	The component of the nonlinear conductivity tensor we are interested in is  $\sigma^{(2)}_{xxx}$, since the system is one dimensional and we only consider a polarization of the light parallel to chain.
	
	\begin{figure*}[ht]
		\centering
		\includegraphics[width=1.0\linewidth]{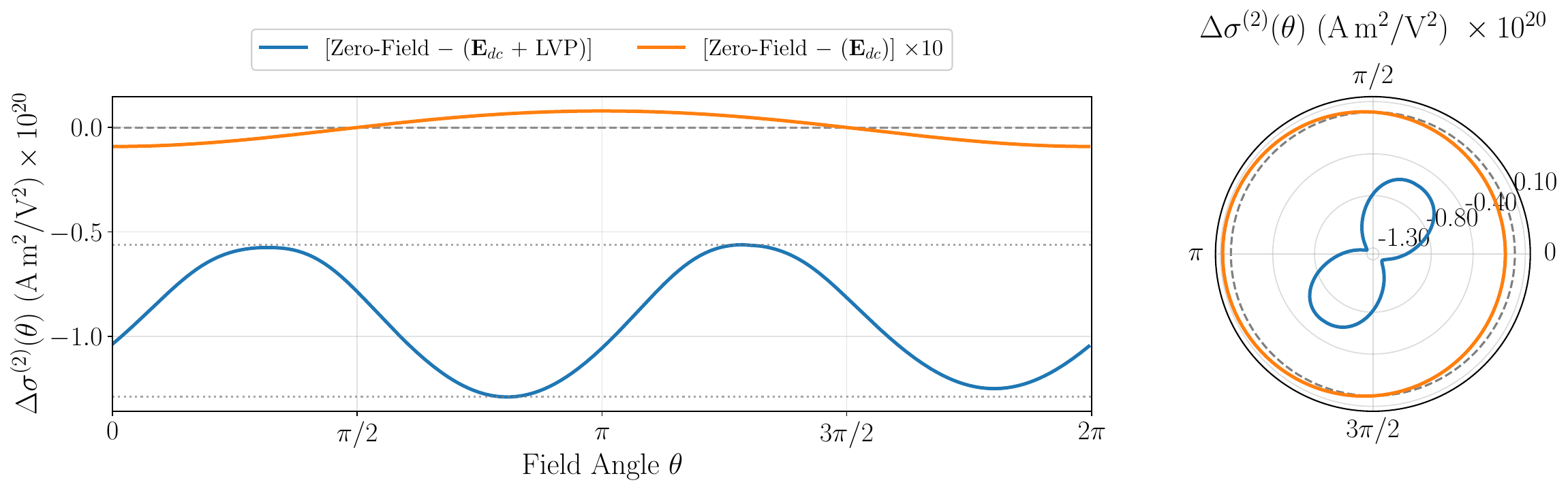}
		\caption{
			Angular dependence of the shift conductivity at $\hbar\omega=E_{\mathrm{gap}}$.
			(left) Relative conductivity $\Delta\sigma(\theta,\hbar\omega=E_{\mathrm{gap}})$ showing a $\pi$-periodic modulation in the presence of the LVP (blue), in contrast with the weakly $2\pi$-periodic response of the pure static field (orange, scaled by $\times 10$ for visibility of the amplitude). 
			(right) Polar representation of the same quantity $\Delta\sigma(\theta, E_{\mathrm{gap}})$, highlighting the angular modulation of the photocurrent.
		}
		\label{fig:panel}
	\end{figure*}
	
	Because the LVP is odd in $k$ and enters the inter-sublattice hopping sector through $\tau_y$, it reshapes the complex phase of the Bloch spinors and therefore the phase of the interband dipole matrix elements. This generates a $k$-odd contribution to the shift vector $R^a_{mn}(k)$ and, consequently, a field-odd contribution to the shift conductivity. The angular dependence is governed by the factor $(\mathbf v\times\mathbf E)_x$ together with the associated spin contribution, both of which vary continuously with the field orientation $\theta$. As a result, the LVP introduces a new angular component in $\sigma^{(2)}(\omega,\theta)$, leading to a $\pi$-periodic modulation that is absent in the purely static-field case, where the response remains weakly $2\pi$ periodic. This change in periodicity constitutes the primary signature of the LV background in the nonlinear optical response. Details of the gauge fixing, numerical stabilization, and implementation are given in Supplementary Material S2.
	
	
	\textit{Results ---}
	The calculations use a one-dimensional $k$ mesh with $N_k=5001$, a Gaussian broadening $\eta=0.015 \mathrm{eV}$, and a field amplitude $E_0=10^{5}\,\mathrm{V\,m^{-1}}$ ($=1\,\mathrm{kV\,cm^{-1}}$), well within experimentally accessible strengths \cite{wang_giant_2024,hiraoka_terahertz_2025,leisgang_giant_2020,klein_electric-field_2017,wang_electrically_2022}, while The LV coupling parameter is set to $\xi=10^{-24}\,\mathrm{C\,m}$. Since the unperturbed bands are spin degenerate, a small Zeeman splitting is introduced to stabilize the gauge, as discussed in Supplementary Material S2. 
	The corresponding shift conductivity for two field orientations separated by $\pi/2$ is shown in Fig.~\ref{fig:linear}. The solid black curve represents the zero-field case, where both the static contribution $H_{\mathrm{ext}}$ and the LV perturbation $H_{\mathrm{LV}}$ are absent, so that the response is determined solely by the Rice--Mele model.
	For a non-zero static field, in the Lorentz-symmetric case (without LVP), the spectra for $\theta$ and $\theta+\pi/2$ nearly coincide, indicating a weak dependence on the field orientation. In contrast, when the LVP is included, the response deviates from the zero-field baseline.
	
	The angular dependence at $\hbar\omega=E_\mathrm{gap}$ is shown in Fig.~\ref{fig:panel}, where we plot the difference with respect to the zero-field response. In the absence of the LV perturbation, the static field contribution $H_{\mathrm{ext}}$ produces only a weak and essentially $2\pi$-periodic modulation [Fig. \ref{fig:panel}(left)], which vanishes when the field is perpendicular to the transverse sublattice displacement (i.e., at $\theta=\pi/2$ and $3\pi/2$). 
	Once the LV contribution $H_{\mathrm{LV}}(k)$ is included, the response develops a pronounced angular modulation with a periodicity of $\pi$, in sharp contrast with the static-field case. Moreover, the gray dotted lines that indicate the extrema of the difference with LVP make it clearer that the peak intensities differ, thereby illustrating the impact of the $k$-odd feature of LVP. Overall, the LVP changes not only the intensity of the shift conductivity, but it does with a clear angular dependency.
	For the arbitrary choice of a generic $\phi=\beta=\pi/4$, the angular profile is shifted away from the reference directions $\theta=0,\ \pi/2,\ \pi,$ and $3\pi/2$. The extrema of the LV-induced response occur near $\theta\approx 55^\circ$ and $\theta\approx 55^\circ+\pi$ (minima), and $\theta\approx 55^\circ+\pi/2$ and $\theta\approx 55^\circ+3\pi/2$ (maxima) [Fig.~\ref{fig:panel}(right)]. This phase offset is not universal, as it depends on the orientation of the background vector $\mathbf v$, but the underlying $\pi$-periodic modulation remains robust.
	
	The origin of this behavior can be traced directly to the structure of the LVP in the 1D model. In the present geometry, rotating $\theta$ modifies the coupling between the applied field and the background vector through the factor $(\mathbf v\times\mathbf E(\theta))_x$, while the spin sector follows the same angular dependence via $(\sin\theta\,\sigma_y+\cos\theta\,\sigma_z)$. This introduces a $k$-odd correction to the Bloch Hamiltonian whose sign and magnitude vary with $\theta$, leading to an asymmetric modification of the interband phase between positive and negative crystal momenta. As a result, the shift vector acquires an additional angular component, giving rise to the observed $\pi$-periodic modulation in the nonlinear response, in contrast with the $2\pi$-periodic behavior of the purely static-field case shown in Fig. \ref{fig:panel}.
	
	To estimate the measurable signal associated with the LV-induced modification of the angular response, we express the dc photocurrent in terms of the optical intensity at the sample. For a monochromatic drive, the photocurrent scales as $I(\theta)\sim \sigma^{(2)}(\omega,\theta)\,|E(\omega)|^2$,
	so that the LV-induced current contrast $\delta I(\theta)\equiv I_{\rm LV}(\theta)-I_{0}(\theta)$ is directly related to the corresponding conductivity difference,
	\begin{equation}
		\delta I(\theta)\simeq \delta \sigma^{(2)}(\theta)\,
		\frac{2I_{\rm opt}}{nc\varepsilon_0},
	\end{equation}
	where $\delta \sigma^{(2)}(\theta)\equiv \sigma^{(2)}_{\rm LV}(\omega,\theta)-\sigma^{(2)}_{0}(\omega,\theta) $
	and $I_{\rm opt}=\tfrac{1}{2}nc\varepsilon_0|E(\omega)|^2$ is the optical intensity. Since the calculated shift conductivity is expressed in units of $\mathrm{A\,m^2/V^2}$, this relation yields the photocurrent directly within the present effective one-dimensional units of the system.
	Using a representative conductivity contrast at $\hbar\omega=E_{\mathrm{gap}}$ of order $\delta \sigma^{(2)}\sim (0.8\text{--}1.1)\times10^{-20}\,\mathrm{A\,m^{2}/V^{2}}$ and a focused optical intensity $I_{\rm opt}\sim 10^{4}\,\mathrm{W/cm^{2}}$ \cite{krishna_understanding_2023,zhang_enhanced_2019}, we obtain a current contrast in the range $\delta I\sim (6\text{--}8)\times10^{-10}\,\mathrm{A}$, i.e. in the range of nA for $n\sim1$, within the resolution of standard low-noise photocurrent measurements \cite{xue_ws2_2024}.
	
	More importantly, the central experimental signature is not simply the magnitude of the photocurrent, but the structure of its angular dependence. While the purely static-field response remains dominated by a single $2\pi$-periodic component, the inclusion of the LVP modifies the angular dependence of the signal, leading into a $\pi$-periodic pattern. 
	This change in periodicity is independent of the overall signal amplitude and persists for generic orientations of the background vector $\mathbf v$, making it an angle-resolved and therefore stable signature of the perturbation.
	This distinction can be identified directly through Fourier analysis of the measured $I(\theta)$ or $\sigma^{(2)}(\theta)$ over several rotations of the static field, where the Lorentz-symmetric case is characterized by a dominant fundamental $2\pi$-periodic component, whereas the LV-induced response exhibits a strong component at twice the angular frequency. Within this framework, the peak-to-valley variation determines the overall magnitude of the signal, while the angular dependence of the response provides a signature of the underlying LVP perturbation.
	
	Although the minimal theory is formulated for a single RC chain, a more realistic device would consist of an aligned bundle or array of weakly interacting polar chains contacted in parallel, so that the individual photocurrent contributions add and enhance the measurable signal without altering the underlying symmetry mechanism. The RC model should therefore be viewed as an effective description of the microscopic building block, while the experimentally accessible current is naturally amplified at the device level. Because the LV-induced $\pi$-periodic component scales linearly with the perturbation strength, a resolvable change in the angular Fourier spectrum can be translated into a sensitivity to the coupling $\xi$. For the parameters used here, with $E_0=10^{5}\,\mathrm{V/m}$, resolving a dominant $\pi$-periodic contribution corresponds to a detectable coupling strength of order $\xi\sim10^{-24}\,\mathrm{C\,m}$ \cite{xue_ws2_2024}. This should be understood as the smallest value of $\xi$ that could be detected with this single atomic chain setup, rather than as a fundamental bound Lorentz-violating coefficients.
	
	\textit{Conclusion ---}
	We have proposed an angular-resolved method to probe LV backgrounds in noncentrosymmetric crystalline systems through the nonlinear shift photocurrent. Using a spinful Rice--Mele model, we showed that an LV background generates a momentum-odd correction to the Bloch Hamiltonian, which reshapes the phase of the interband dipole matrix elements and modifies the angular structure of the nonlinear response.
	In contrast to the weakly $2\pi$-periodic behavior obtained from the purely static-field contribution, the LV perturbation induces a robust $\pi$-periodic modulation in the shift conductivity $\sigma^{(2)}(\theta)$, with extrema separated by $\pi/2$. While the phase offset of this modulation depends on the orientation of the background vector $\mathbf{v}$, its angular dependence is generic and provides a clear signature of the LV background.
	Crucially, the LV signal can be identified through Fourier analysis of the angular response, where the emergence of a dominant $\pi$-periodic component replaces the $2\pi$ baseline of the Lorentz-symmetric case. This angle-resolved detection scheme avoids reliance on absolute current magnitudes and offers a robust experimental route to isolating LV-induced effects.
	For the parameters considered here, resolving this change in angular periodicity corresponds to a sensitivity to the LV coupling of order $\xi\sim10^{-24}\,\mathrm{C\,m}$. This value should be interpreted as the smallest coupling strength that could be detected in this system, rather than a fundamental limit. Our results establish angular-resolved nonlinear optical transport as a practical probe of Lorentz-violating effects in atomic chain polar systems.
	
	\textit{Acknowledgments ---} The authors acknowledge financial support from the Brazilian funding agencies FAPES (1044/2022, 1081/2022 - P:2022-8L35F, and 875/2023 - P:2023-V36VC), and CNPq (under grants 444450/2024-6, 305227/2024-6 and 442781/2023-7) and are also grateful for the computational resources of the Sci-Com Lab/UFES. We also acknowledge financial support from MICIU/AEI/10.13039/501100011033 through Grants TED2021-131323B-I00 and PID2022-141712NB-C21, the María de Maeztu Program for Units of Excellence in R\&D through Grant CEX2023-001316-M, the Comunidad de Madrid within the Recovery, Transformation and Resilience Plan through the project “Disruptive 2D materials” (MAD2D-CM-UAM7) funded by the NextGenerationEU programme from the European Union, and we also thank the Naturgy Foundation. We also acknowledge computational resources from CCC at UAM and Spanish RES.
	
\bibliographystyle{ieeetr}
\bibliography{reference}	
\clearpage
\onecolumngrid
\include{SI}

\end{document}

%% file: SI.tex
\hypersetup{colorlinks=true, linkcolor=blue, urlcolor=blue, citecolor=blue}
\renewcommand{\thefigure}{S\arabic{figure}}
\setcounter{figure}{0} 

\setlength{\parindent}{0pt}


\section*{Supplemental Material for:\\
``Nonlinear optical response as a probe of emergent Lorentz symmetry violation in noncentrosymmetric materials"}

\author{Guilherme J. Inacio}
\affiliation{Departamento de Física, Universidade Federal do Espírito Santo, 29075-910 Vitória-ES, Brazil}
\affiliation{Departamento de F\'isica de la Materia Condensada, Universidad Aut\'{o}noma de Madrid, E-28049 Madrid, Spain}

\author{Nathanael N. Batista}
\affiliation{Departamento de Física, Universidade Federal do Espírito Santo, 29075-910 Vitória-ES, Brazil}

\author{Wesley Spalenza}
\affiliation{Instituto Federal do Esp\'{i}rito Santo, IFES - Cariacica, ES, Brasil}
\affiliation{Centro Brasileiro de Pesquisas Físicas, CBPF, RJ, Brasil}

\author{Humberto Belich}
\affiliation{Departamento de Física, Universidade Federal do Espírito Santo, 29075-910 Vitória-ES, Brazil}

\author{Juan Jos\'e Palacios}
\affiliation{Departamento de F\'isica de la Materia Condensada, Universidad Aut\'{o}noma de Madrid, E-28049 Madrid, Spain}
\affiliation{Condensed Matter Physics Center (IFIMAC), Universidad Aut\'{o}noma de Madrid, E-28049 Madrid, Spain}
\affiliation{Instituto Nicol\'as Cabrera (INC), Universidad Aut\'{o}noma de Madrid, E-28049 Madrid, Spain}

\author{Wendel S. Paz}
\email{wendel.paz@ufes.br, corresponding author}
\affiliation{Departamento de Física, Universidade Federal do Espírito Santo, 29075-910 Vitória-ES, Brazil}
\affiliation{Departamento de F\'isica de la Materia Condensada, Universidad Aut\'{o}noma de Madrid, E-28049 Madrid, Spain}

\date{\today}

\maketitle

\section{S1. Derivation of the Lorentz-Violating Hamiltonian Correction}
We outline the main steps of the derivation and retain only the terms
relevant for the effective low-energy Hamiltonian used in the main text.
We begin with the gauge-invariant Dirac equation incorporating a non-minimal Lorentz-violating (LV) coupling:
\begin{equation}
    (i c \hbar \gamma ^{\mu }{\cal D}_{\mu } - m c^2)\Psi = 0,  \label{eq:dirac}
\end{equation}
with the extended covariant derivative:
\begin{equation}
    {\cal D}_{\mu} = \partial_{\mu} + i \frac{e}{\hbar c} A_{\mu} + i \frac{g}{\hbar c} v^{\nu} \tilde{F}_{\mu \nu},
\end{equation}
where $\tilde{F}^{\mu\nu} = \frac{1}{2} \epsilon^{\mu\nu\alpha\beta} F_{\alpha\beta}$ is the dual electromagnetic tensor and $v^\mu = (v^0, \mathbf{v})$ is a fixed background four-vector that breaks Lorentz symmetry.

\subsection*{2. Dirac Spinor Decomposition}

To obtain the non-relativistic limit, we write the Dirac spinor as:
\[
\Psi = 
\begin{pmatrix}
\phi \\
\chi
\end{pmatrix},
\]
where $\phi$ and $\chi$ are the large and small components, respectively.

Using the Dirac gamma matrices in the Dirac representation:
\[
\gamma^0 = 
\begin{pmatrix}
\mathbb{I} & 0 \\
0 & -\mathbb{I}
\end{pmatrix}, \quad
\gamma^i = 
\begin{pmatrix}
0 & \sigma^i \\
-\sigma^i & 0
\end{pmatrix},
\]
we can write the Dirac equation as a coupled system:
\begin{align}
\left( E - eV - g\,\mathbf{v} \cdot \mathbf{B} \right)\, \phi 
&= \boldsymbol{\sigma} \cdot 
\Big[ \mathbf{p} - \frac{e}{c}\mathbf{A} 
+ g v^0 \mathbf{B} - g\,\mathbf{v} \times \frac{\mathbf{E}}{c} \Big]\, \chi 
+ m c^2\, \phi, \\
\left( E - eV - g\,\mathbf{v} \cdot \mathbf{B} \right)\, \chi 
&= \boldsymbol{\sigma} \cdot 
\Big[ \mathbf{p} - \frac{e}{c}\mathbf{A} 
+ g v^0 \mathbf{B} - g\,\mathbf{v} \times \frac{\mathbf{E}}{c} \Big]\, \phi 
- m c^2\, \chi.
\end{align}

\subsection*{3. Foldy–Wouthuysen Expansion}

In the non-relativistic limit ($E \approx m c^2 + \varepsilon$, with $\varepsilon \ll m c^2$), the small component can be approximated as:
\begin{equation}
    \chi \approx \frac{1}{2mc} \, \boldsymbol{\sigma} \cdot \left( \mathbf{p} - \frac{e}{c} \mathbf{A} + g v^0 \mathbf{B} - g \, \mathbf{v} \times \frac{\mathbf{E}}{c} \right) \phi.
\label{eq:chi}
\end{equation}

Substituting Eq.~\eqref{eq:chi} into the upper equation yields the effective Schrödinger-Pauli equation for $\phi$:
\begin{equation}
    \left( E - m c^2 - e V - g \mathbf{v} \cdot \mathbf{B} \right)\phi 
    = \frac{1}{2m} \left( \boldsymbol{\sigma} \cdot \boldsymbol{\Pi} \right)^2 \phi,
\end{equation}
where the generalized momentum is
\begin{equation}
    \boldsymbol{\Pi}
    =\underbrace{(\mathbf{p} - e\mathbf{A})}_{\boldsymbol{\Pi}_{\rm em}}
    +\underbrace{(g v_0\mathbf{B} - g\,\mathbf{v}\times\mathbf{E})}_{\boldsymbol{\Delta}}.
    \label{eq:SM_Pi_def}
\end{equation}
Throughout this section we set $\hbar=c=1$. \\

Using the Pauli identity
\[
    (\boldsymbol{\sigma}\!\cdot\!\boldsymbol{\Pi})^2
    =\boldsymbol{\Pi}^{\,2}+\frac{i}{2}\,\epsilon_{abc}\sigma_c[\Pi_a,\Pi_b],
\]
and computing the commutator
\begin{align}
    [\Pi_a,\Pi_b]
    &=[\Pi_{{\rm em},a},\Pi_{{\rm em},b}]
    +[\Pi_{{\rm em},a},\Delta_b]
    +[\Delta_a,\Pi_{{\rm em},b}]
    \nonumber\\
    &=i\,\epsilon_{abk}\big[eB_k+(\boldsymbol{\nabla}\times\boldsymbol{\Delta})_k\big],
\end{align}
we obtain
\begin{equation}
    (\boldsymbol{\sigma}\!\cdot\!\boldsymbol{\Pi})^2
    =\boldsymbol{\Pi}^{\,2}
    -\boldsymbol{\sigma}\!\cdot\!\big[e\,\mathbf{B}+\boldsymbol{\nabla}\times\boldsymbol{\Delta}\big].
    \label{eq:SM_Pauli_identity}
\end{equation}

Substituting Eq.~\eqref{eq:SM_Pauli_identity} into the Pauli equation yields the effective Hamiltonian
\begin{equation}
    H
    =eA_0
    +\frac{1}{2m_e}\,\boldsymbol{\Pi}^{\,2}
    -\frac{1}{2m_e}\,\boldsymbol{\sigma}\!\cdot\!\big[e\,\mathbf{B}+\boldsymbol{\nabla}\times\boldsymbol{\Delta}\big].
    \label{eq:SM_H_before_expand}
\end{equation}
\\

Using $\boldsymbol{\Pi}=\boldsymbol{\Pi}_{\rm em}+\boldsymbol{\Delta}$ and
$(\mathbf{a}+\mathbf{b})^2=\mathbf{a}^{\,2}+\{\mathbf{a},\mathbf{b}\}+\mathbf{b}^{\,2}$,
\begin{equation}
    H
    =eA_0
    +\frac{1}{2m_e}(\mathbf{p}-e\mathbf{A})^2
    -\frac{e}{2m_e}\,\boldsymbol{\sigma}\!\cdot\!\mathbf{B}
    +\frac{1}{2m_e}\{\boldsymbol{\Pi}_{\rm em},\boldsymbol{\Delta}\}
    +\frac{1}{2m_e}\,\boldsymbol{\Delta}^{\,2}
    -\frac{1}{2m_e}\,\boldsymbol{\sigma}\!\cdot\!(\boldsymbol{\nabla}\times\boldsymbol{\Delta}).
    \label{eq:SM_H_expand}
\end{equation}

For weakly varying fields, $[\mathbf{p},\boldsymbol{\Delta}]\simeq -i\boldsymbol{\nabla}\boldsymbol{\Delta}$,
and taking $\boldsymbol{\nabla}\!\cdot\!\boldsymbol{\Delta}\simeq0$, the mixed anticommutator gives
\begin{equation}
    \frac{1}{2m_e}\{\boldsymbol{\Pi}_{\rm em},\boldsymbol{\Delta}\}
    \simeq
    \frac{1}{m_e}\,\boldsymbol{\Delta}\!\cdot\!(\mathbf{p}-e\mathbf{A}).
    \label{eq:SM_anticommutator}
\end{equation}

The quadratic and curl terms follow from
\begin{align}
    \boldsymbol{\Delta}^{\,2}
    &=(g v_0\mathbf{B} - g\,\mathbf{v}\times\mathbf{E})^2
    =g^2\!\left[(v_0)^2\mathbf{B}^{\,2}
    -2v_0\,\mathbf{B}\!\cdot\!(\mathbf{v}\times\mathbf{E})
    +|\mathbf{v}\times\mathbf{E}|^2\right],
    \label{eq:SM_Delta_sq}
    \\
    \boldsymbol{\nabla}\times\boldsymbol{\Delta}
    &=g v_0(\boldsymbol{\nabla}\times\mathbf{B})
    - g\,\boldsymbol{\nabla}\times(\mathbf{v}\times\mathbf{E}).
    \label{eq:SM_curl}
\end{align}

Substituting Eqs.~\eqref{eq:SM_anticommutator}–\eqref{eq:SM_curl} into
Eq.~\eqref{eq:SM_H_expand} and discarding the $(v_0)^2\mathbf{B}^{\,2}$ term (second order in $\mathbf{B}$), and $g^2$ term,
we obtain
\begin{align}
    H
    =& 
    \Bigg[
    \frac{1}{2m_e}(\mathbf{p}-e\mathbf{A})^2
    + eA_0
    - \frac{e}{2m_e}\,(\boldsymbol{\sigma}\!\cdot\!\mathbf{B})
    \Bigg]
    & + \frac{g v_0}{2m_e}\,\boldsymbol{\sigma}\!\cdot\!(\boldsymbol{\nabla}\times\mathbf{B})
     - \frac{g}{2m_e}\,\boldsymbol{\sigma}\!\cdot\!\big[\boldsymbol{\nabla}\times(\mathbf{v}\times\mathbf{E})\big]
    \notag\\[4pt]
    & + \frac{g v_0}{m_e}\,(\mathbf{p}-e\mathbf{A})\!\cdot\!\mathbf{B}
     - \frac{g}{m_e}\,(\mathbf{p}-e\mathbf{A})\!\cdot\!(\mathbf{v}\times\mathbf{E}) \, .
    \label{eq:SM_H_linear_g}
\end{align}

Setting $\mathbf A=\mathbf 0$ and $\mathbf B=\mathbf 0$ in Eq.~\eqref{eq:SM_H_linear_g}, the nonrelativistic Hamiltonian reduces to
\begin{equation}
    \label{eq:H_AB0_step}
    H=\frac{\mathbf p^2}{2m_e}+eA_0
    -\frac{g}{2m_e}\,\boldsymbol{\sigma}\!\cdot\!\big[\boldsymbol{\nabla}\times(\mathbf v\times\mathbf E)\big]
    -\frac{g}{m_e}\,\mathbf p\!\cdot\!(\mathbf v\times\mathbf E),
\end{equation}
from which we retain the two terms linear in $g$: the spin contribution and the orbital one. Assuming constant $\mathbf v$ and charge neutrality, $\boldsymbol{\nabla}\!\cdot\!\mathbf E=0$, one has
\begin{equation}
    \boldsymbol{\nabla}\times(\mathbf v\times\mathbf E)=-(\mathbf v\!\cdot\!\boldsymbol{\nabla})\mathbf E
      \longrightarrow  
    -i(\mathbf v\!\cdot\!\mathbf k)\mathbf E
\end{equation}
in momentum space. 
For a one-dimensional chain along $\hat{\mathbf x}$, we take
\begin{align}
    \mathbf k &= k\,\hat{\mathbf x},\\
    \mathbf v &= v_0\big(\sin\beta\cos\phi\,\hat{\mathbf x}
    +\sin\beta\sin\phi\,\hat{\mathbf y}
    +\cos\beta\,\hat{\mathbf z}\big),\\
    \mathbf E(\theta) &= E_0\big(\sin\theta\,\hat{\mathbf y}
    +\cos\theta\,\hat{\mathbf z}\big),
    \label{eq:geom_choice}
\end{align}
so that
\begin{equation}
    \mathbf v\!\cdot\!\mathbf k=v_0\sin\beta\cos\phi\,k,
    \qquad
    \boldsymbol{\sigma}\!\cdot\!\mathbf E
    =E_0(\sin\theta\,\sigma_y+\cos\theta\,\sigma_z).
\end{equation}

The ordinary electrostatic coupling of the rotated static field to the two sublattices gives a $k$-independent contribution,
\begin{equation}
    \label{eq:Hdc_SM}
    H_{\mathrm{dc}}
    =
    \delta U(\theta)\,\tau_z\otimes\sigma_0,
\end{equation}
where $\delta U(\theta)=(U_A-U_B)/2$ is the staggered onsite shift generated by the field. In terms of the transverse sublattice coordinates $(y_A,z_A)$ and $(y_B,z_B)$,
\begin{equation}
    U_\alpha(\theta)=-e\,\mathbf E(\theta)\!\cdot\!\mathbf r_\alpha
    =-eE_0\big(y_\alpha\sin\theta+z_\alpha\cos\theta\big),
    \qquad \alpha=A,B,
\end{equation}
so that, after discarding the uniform onsite shift,
\begin{equation}
    \label{eq:deltaU_SM}
    \delta U(\theta)
    =
    -\frac{eE_0}{2}
    \Big[(y_A-y_B)\sin\theta+(z_A-z_B)\cos\theta\Big].
\end{equation}
This term defines the purely static-field baseline used in the main text.

The LV spin term follows from Eq.~\eqref{eq:H_AB0_step} as
\begin{equation}
    \delta H_{\mathrm{LV}}^{(\mathrm{spin})}(k)
    =
    \frac{i g}{2m_e}\,
    (\mathbf v\!\cdot\!\mathbf k)\,
    \boldsymbol{\sigma}\!\cdot\!\mathbf E
    =
    \frac{i g}{2m_e}\,
    v_0E_0\sin\beta\cos\phi\,k\,
    (\sin\theta\,\sigma_y+\cos\theta\,\sigma_z).
\end{equation}
The factor of $i$ reflects the momentum-space representation of the gradient operator and is absorbed into the effective lattice representation, where Hermiticity is ensured by the $\tau_y$ structure. 
In the effective two-sublattice description, its odd-in-$k$ structure is encoded through $\tau_y$, and after restoring $\hbar$ and using $g=\xi/c$ we write
\begin{equation}
    \delta H_{\mathrm{LV}}^{(\mathrm{spin})}(k)
    =
    \lambda_{\mathrm{spin}}(k)\,
    \tau_y\otimes(\sin\theta\,\sigma_y+\cos\theta\,\sigma_z),
    \qquad
    \lambda_{\mathrm{spin}}(k)=
    \frac{\hbar\xi E_0}{2m_ec}\,
    v_0\sin\beta\cos\phi\,k.
    \label{eq:spin_SM_final}
\end{equation}

The orbital term is
\begin{equation}
    \delta H_{\mathrm{LV}}^{(\mathrm{orb})}
    =
    -\frac{g}{m_e}\,\mathbf p\!\cdot\!(\mathbf v\times\mathbf E).
\end{equation}
For the geometry of Eq.~\eqref{eq:geom_choice},
\begin{equation}
    (\mathbf v\times\mathbf E)_x
    =
    E_0\big(v_y\cos\theta-v_z\sin\theta\big)
    =
    E_0v_0\big(\sin\beta\sin\phi\cos\theta-\cos\beta\sin\theta\big),
\end{equation}
and since only $p_x=\hbar k$ contributes in 1D,
\begin{equation}
    \label{eq:orb_SM_final}
    \delta H_{\mathrm{LV}}^{(\mathrm{orb})}(k)
    =
    \lambda_{\mathrm{orb}}(k)\,\tau_y\otimes\sigma_0,
    \qquad
    \lambda_{\mathrm{orb}}(k)
    =
    -\frac{\hbar\xi E_0}{m_ec}\,
    v_0\big(\sin\beta\sin\phi\cos\theta-\cos\beta\sin\theta\big)\,k.
\end{equation}

Combining both pieces, the Lorentz-violating correction in the effective 1D model is
\begin{equation}
    H_{\mathrm{LV}}(k)
    =
    \tau_y\otimes
    \Big[
    \lambda_{\mathrm{orb}}(k)\,\sigma_0
    +
    \lambda_{\mathrm{spin}}(k)\,
    (\sin\theta\,\sigma_y+\cos\theta\,\sigma_z)
    \Big].
    \label{eq:H_LV_1D_final}
\end{equation}
The full model used in the calculations is therefore
\begin{equation}
    H_{\mathrm{eff}}(k)
    =
    H_{\mathrm{RM}}(k)+H_{\mathrm{LV}}(k)+H_{\mathrm{dc}},
    \label{eq:Heff_SM_final}
\end{equation}
where $H_{\mathrm{dc}}$ is even in $k$ and sets the pure static-field response, while $H_{\mathrm{LV}}(k)$ is odd in $k$ and modifies the phase structure of the Bloch states through the $\tau_y$ channel.

\section{S2. Shift-conductivity formalism and numerical implementation}
    
    The interband dipole matrix elements are evaluated in the length gauge at fixed crystal momentum,
    \begin{equation}
        r_{mn}(\mathbf{k}) = i\,\frac{\langle u_m(\mathbf{k})|\partial_{\mathbf{k}} H(\mathbf{k})|u_n(\mathbf{k})\rangle}{E_n(\mathbf{k})-E_m(\mathbf{k})},
    \label{eq:dipole_supp}
    \end{equation}
    where $|u_n(\mathbf{k})\rangle$ are the cell-periodic Bloch eigenstates of the full Hamiltonian and $\partial_{\mathbf{k}} H$ is obtained analytically. The diagonal Berry connections are defined as
    \begin{equation}
        A_{nn}(\mathbf{k}) = i\,\langle u_n(\mathbf{k})|\partial_{\mathbf{k}}u_n(\mathbf{k})\rangle,
    \label{eq:berryconn_supp}
    \end{equation}
    and are computed numerically from overlaps between eigenstates at neighboring $\mathbf{k}$-points.
    
    To ensure numerical smoothness, the phases of the eigenstates are fixed by imposing a continuous gauge across the Brillouin zone. At each $\mathbf{k}$, we extract
    \begin{equation}
        \theta_{n\mathbf{k}}=\arg\!\left[\sum_{\alpha}U_{\alpha n}(\mathbf{k})\right],
    \end{equation}
    and redefine the eigenvectors as
    \begin{equation}
        \tilde{u}_n(\mathbf{k})=e^{-i\theta_{n\mathbf{k}}}u_n(\mathbf{k}),
    \end{equation}
    so that $\sum_{\alpha}\tilde{U}_{\alpha n}(\mathbf{k})$ corresponds to fixing the global phase of each eigenstate by requiring that the sum over its components is real and positive.
    This procedure enforces a smooth gauge along $\mathbf{k}$, stabilizing both the Berry connections and the phase of the dipole matrix elements~\cite{garcia_blazquez_shift_2023,esteve_paredes_excitons_2025}.
    
    Writing
    \begin{align}
        r_{mn}(\mathbf{k}) = |r_{mn}(\mathbf{k})|\,e^{i\phi_{mn}(\mathbf{k})},
    \end{align}
    the shift conductivity is given in the length gauge by~\cite{sipe2000second}
    \begin{equation}
        \sigma^{(2)}_{abb}(\omega,\theta)
        = -\frac{\pi e^{3}}{\hbar^{2}}
        \sum_{m\neq n}\int\frac{dk}{2\pi}\,
        f_{mn}(k)\,|r^{b}_{mn}(k)|^{2}\,
        R^{a}_{mn}(k)\,
        \delta\big(E_{mn}(k)-\omega\big),
    \label{eq:shift_general_supp}
    \end{equation}
    In this expression, $f_{mn}(k)=f_m(k)-f_n(k)$ is the occupation factor. 
    At zero temperature with a fully occupied valence band and an empty conduction band, $f_{mn}(k)=1$ for all allowed interband transitions, which is the regime used in our numerical evaluation. 
    The parameter sensitive to the perturbation is the shift vector, defined as
    \begin{equation}
        R^{a}_{mn}(\mathbf{k})=\partial_{k_a}\phi_{mn}(\mathbf{k})
        -\big(A^{a}_{mm}(\mathbf{k})-A^{a}_{nn}(\mathbf{k})\big)
    \label{eq:shift_vector_supp}
    \end{equation}
    The coupling to the Lorentz-violating term is controlled by the projection $(\mathbf{v}\times\mathbf{E}(\theta))_x$, which depends on the transverse components of the applied field. This angular dependence induces a $k$-odd modification of the interband phase and therefore alters the shift vector $R^a_{mn}(k)$. The phase derivative is evaluated numerically using a symmetric finite-difference form,
    \begin{equation}
        \partial_{k}\phi_{mn}(\mathbf{k})
        \simeq \frac{1}{2\,dk}\,
        \arg\!\big[r_{mn}(\mathbf{k}+dk)\,r_{mn}(\mathbf{k}-dk)^*\big],
        \end{equation}
    which depends only on phase differences and is therefore gauge-invariant.


%% file: reference.bib
@article{pietralonga2026probing,
  title={Probing shift-current responses in noncentrosymmetric materials using quantum algorithms},
  author={Pietralonga, Cleoner S and Paz, Wendel S},
  journal={Materials Today Quantum},
  pages={100066},
  year={2026},
  publisher={Elsevier}
}

@article{wang_giant_2024,
	title = {Giant electric field-induced second harmonic generation in polar skyrmions},
	volume = {15},
	copyright = {2024 The Author(s)},
	issn = {2041-1723},
	url = {https://www.nature.com/articles/s41467-024-45755-5},
	doi = {10.1038/s41467-024-45755-5},
	language = {en},
	number = {1},
	urldate = {2025-11-19},
	journal = {Nature Communications},
	author = {Wang, Sixu and Li, Wei and Deng, Chenguang and Hong, Zijian and Gao, Han-Bin and Li, Xiaolong and Gu, Yueliang and Zheng, Qiang and Wu, Yongjun and Evans, Paul G. and Li, Jing-Feng and Nan, Ce-Wen and Li, Qian},
	month = feb,
	year = {2024},
	note = {Publisher: Nature Publishing Group},
	pages = {1374},
}

@article{hiraoka_terahertz_2025,
	title = {Terahertz field effect in a two-dimensional semiconductor},
	volume = {16},
	copyright = {2025 The Author(s)},
	issn = {2041-1723},
	url = {https://www.nature.com/articles/s41467-025-60588-6},
	doi = {10.1038/s41467-025-60588-6},
	language = {en},
	number = {1},
	urldate = {2025-11-18},
	journal = {Nature Communications},
	author = {Hiraoka, Tomoki and Nestler, Sandra and Zhang, Wentao and Rossel, Simon and Hafez, Hassan A. and Fabretti, Savio and Schlörb, Heike and Thomas, Andy and Turchinovich, Dmitry},
	month = jun,
	year = {2025},
	note = {Publisher: Nature Publishing Group},
	pages = {5235},
}

@article{wang_electrically_2022,
	title = {Electrically {Tunable} {Second} {Harmonic} {Generation} in {Atomically} {Thin} {ReS2}},
	volume = {16},
	issn = {1936-0851},
	url = {https://doi.org/10.1021/acsnano.2c00514},
	doi = {10.1021/acsnano.2c00514},
	number = {4},
	urldate = {2025-07-28},
	journal = {ACS Nano},
	author = {Wang, Jing and Han, Nannan and Luo, Zheng-Dong and Zhang, Mingwen and Chen, Xiaoqing and Liu, Yan and Hao, Yue and Zhao, Jianlin and Gan, Xuetao},
	month = apr,
	year = {2022},
	note = {Publisher: American Chemical Society},
	pages = {6404--6413},
}

@article{klein_electric-field_2017,
	title = {Electric-{Field} {Switchable} {Second}-{Harmonic} {Generation} in {Bilayer} {MoS2} by {Inversion} {Symmetry} {Breaking}},
	volume = {17},
	issn = {1530-6984},
	url = {https://doi.org/10.1021/acs.nanolett.6b04344},
	doi = {10.1021/acs.nanolett.6b04344},
	number = {1},
	urldate = {2025-07-28},
	journal = {Nano Letters},
	author = {Klein, J. and Wierzbowski, J. and Steinhoff, A. and Florian, M. and Rösner, M. and Heimbach, F. and Müller, K. and Jahnke, F. and Wehling, T. O. and Finley, J. J. and Kaniber, M.},
	month = jan,
	year = {2017},
	note = {Publisher: American Chemical Society},
	pages = {392--398},
}

@article{leisgang_giant_2020,
	title = {Giant {Stark} splitting of an exciton in bilayer {MoS2}},
	volume = {15},
	copyright = {2020 The Author(s), under exclusive licence to Springer Nature Limited},
	issn = {1748-3395},
	url = {https://www.nature.com/articles/s41565-020-0750-1},
	doi = {10.1038/s41565-020-0750-1},
	language = {en},
	number = {11},
	urldate = {2025-07-31},
	journal = {Nature Nanotechnology},
	author = {Leisgang, Nadine and Shree, Shivangi and Paradisanos, Ioannis and Sponfeldner, Lukas and Robert, Cedric and Lagarde, Delphine and Balocchi, Andrea and Watanabe, Kenji and Taniguchi, Takashi and Marie, Xavier and Warburton, Richard J. and Gerber, Iann C. and Urbaszek, Bernhard},
	month = nov,
	year = {2020},
	note = {Publisher: Nature Publishing Group},
	pages = {901--907},
}

@article{bakke2015influence,
  title={On the influence of a Rashba-type coupling induced by Lorentz-violating effects on a Landau system for a neutral particle},
  author={Bakke, K and Belich, H},
  journal={Annals of Physics},
  volume={354},
  pages={1--9},
  year={2015},
  publisher={Elsevier}
}

@article{bakke2014rashba,
  title={Rashba coupling induced by Lorentz symmetry breaking effects},
  author={Bakke, Knut and Belich, Humberto},
  journal={Annalen der Physik},
  volume={526},
  number={3-4},
  pages={187--194},
  year={2014},
  publisher={Wiley Online Library}
}

@article{rice1982elementary,
  title={Elementary excitations of a linearly conjugated diatomic polymer},
  author={Rice, MJ and Mele, EJ},
  journal={Physical Review Letters},
  volume={49},
  number={19},
  pages={1455},
  year={1982},
  publisher={APS}
}

@article{young2012first,
  title={First principles calculation of the shift current photovoltaic effect in ferroelectrics},
  author={Young, Steve M and Rappe, Andrew M},
  journal={Physical review letters},
  volume={109},
  number={11},
  pages={116601},
  year={2012},
  publisher={APS}
}

@article{Colladay1997,
  author = {D. Colladay and V. A. Kostelecký},
  title = {CPT violation and the standard model},
  journal = {Phys. Rev. D},
  volume = {55},
  pages = {6760},
  year = {1997},
  doi = {10.1103/PhysRevD.55.6760}
}

@article{Colladay1998,
  author = {D. Colladay and V. A. Kostelecký},
  title = {Lorentz-violating extension of the standard model},
  journal = {Phys. Rev. D},
  volume = {58},
  pages = {116002},
  year = {1998},
  doi = {10.1103/PhysRevD.58.116002}
}

@article{Grushin2012,
  author = {A. G. Grushin},
  title = {Consequences of a condensed matter realization of Lorentz-violating QED in Weyl semi-metals},
  journal = {Phys. Rev. D},
  volume = {86},
  pages = {045001},
  year = {2012},
  doi = {10.1103/PhysRevD.86.045001}
}

@article{Kostelecky2004,
  author = {V. A. Kostelecký},
  title = {Lorentz and CPT violation in the Standard Model},
  journal = {Phys. Rev. D},
  volume = {69},
  pages = {105009},
  year = {2004},
  doi = {10.1103/PhysRevD.69.105009}
}

@article{belich2005non,
  title={Non-minimal coupling to a Lorentz-violating background and topological implications},
  author={Belich, H and Costa-Soares, T and FerreiraJr, MM and Helay{\"e}l-Neto, JA},
  journal={The European Physical Journal C-Particles and Fields},
  volume={41},
  pages={421--426},
  year={2005},
  publisher={Springer}
}

@article{bakke2015landau,
  title={A Landau-type quantization from a Lorentz symmetry violation background with crossed electric and magnetic fields},
  author={Bakke, K and Belich, H},
  journal={Journal of Physics G: Nuclear and Particle Physics},
  volume={42},
  number={9},
  pages={095001},
  year={2015},
  publisher={IOP Publishing}
}

@article{kostelecky2022lorentz,
  title={Lorentz violation in Dirac and Weyl semimetals},
  author={Kosteleck{\`y}, V Alan and Lehnert, Ralf and McGinnis, Navin and Schreck, Marco and Seradjeh, Babak},
  journal={Physical Review Research},
  volume={4},
  number={2},
  pages={023106},
  year={2022},
  publisher={APS}
}

@article{Morimoto2016,
  author    = {Morimoto, Takahiro and Nagaosa, Naoto},
  title     = {Topological aspects of nonlinear-optical effects},
  journal   = {Science Advances},
  volume    = {2},
  number    = {5},
  pages     = {e1501524},
  year      = {2016},
  publisher = {American Association for the Advancement of Science}
}

@article{esteve_paredes_excitons_2025,
	title = {Excitons in nonlinear optical responses: shift current in {MoS2} and {GeS} monolayers},
	volume = {11},
	issn = {2057-3960},
	shorttitle = {Excitons in nonlinear optical responses},
	url = {https://www.nature.com/articles/s41524-024-01504-2},
	doi = {10.1038/s41524-024-01504-2},
	language = {en},
	number = {1},
	urldate = {2025-02-10},
	journal = {npj Computational Materials},
	author = {Esteve-Paredes, J. J. and García-Blázquez, M. A. and Uría-Álvarez, A. J. and Camarasa-Gómez, M. and Palacios, J. J.},
	month = jan,
	year = {2025},
	pages = {13},
}

@article{garcia_blazquez_shift_2023,
	title = {Shift {Current} with {Gaussian} {Basis} {Sets} and {General} {Prescription} for {Maximally} {Symmetric} {Summations} in the {Irreducible} {Brillouin} {Zone}},
	volume = {19},
	copyright = {https://creativecommons.org/licenses/by/4.0/},
	issn = {1549-9618, 1549-9626},
	url = {https://pubs.acs.org/doi/10.1021/acs.jctc.3c00917},
	doi = {10.1021/acs.jctc.3c00917},
	language = {en},
	number = {24},
	urldate = {2025-06-08},
	journal = {Journal of Chemical Theory and Computation},
	author = {García-Blázquez, M. A. and Esteve-Paredes, J. J. and Uría-Álvarez, A. J. and Palacios, J. J.},
	month = dec,
	year = {2023},
	pages = {9416--9434},
}

@article{Kostelecky1999,
  author    = {Kosteleck\'{y}, V. Alan and Lane, Charles},
  title     = {Nonrelativistic quantum Hamiltonians for Lorentz violation},
  journal   = {Physical Review D},
  volume    = {60},
  issue     = {11},
  pages     = {116010},
  year      = {1999},
  month     = {Nov},
  doi       = {10.1103/PhysRevD.60.116010},
  publisher = {American Physical Society}
}

@article{Kostelecky2011,
  author    = {Kosteleck\'{y}, V. Alan and Russell, Neil},
  title     = {Data tables for Lorentz and CPT violation},
  journal   = {Reviews of Modern Physics},
  volume    = {83},
  issue     = {1},
  pages     = {11--31},
  year      = {2011},
  month     = {Mar},
  doi       = {10.1103/RevModPhys.83.11},
  publisher = {American Physical Society}
}

@article{Gabrielse1999,
  author    = {Gabrielse, G. and Khabbaz, A. and Hall, D. S. and Heimann, C. and Kalinowsky, H. and Jhe, W.},
  title     = {Precision Mass Spectroscopy of the Antiproton and Proton Using Simultaneously Trapped Particles},
  journal   = {Physical Review Letters},
  volume    = {82},
  issue     = {16},
  pages     = {3198--3201},
  year      = {1999},
  month     = {Apr},
  doi       = {10.1103/PhysRevLett.82.3198},
  publisher = {American Physical Society}
}

@article{FermiLAT2009,
  author    = {Abdo, A. A. and others and {The Fermi LAT and Fermi GBM Collaborations}},
  title     = {A limit on the variation of the speed of light arising from quantum gravity effects},
  journal   = {Nature},
  volume    = {462},
  number    = {7271},
  pages     = {331--334},
  year      = {2009},
  doi       = {10.1038/nature08574}
}

@article{IceCube2018,
  author    = {Aartsen, M. G. and others and {The IceCube Collaboration}},
  title     = {Neutrino interferometry for high-precision tests of Lorentz symmetry with IceCube},
  journal   = {Nature Physics},
  volume    = {14},
  number    = {9},
  pages     = {961--966},
  year      = {2018},
  doi       = {10.1038/s41567-018-0172-2}
}

@article{ALPHA2017,
  author    = {Ahmadi, M. and others and {The ALPHA Collaboration}},
  title     = {Observation of the 1S–2S transition in trapped antihydrogen},
  journal   = {Nature},
  volume    = {541},
  number    = {7638},
  pages     = {506--510},
  year      = {2017},
  doi       = {10.1038/nature21040}
}

@article{Muong-2_2008,
  author    = {Bennett, G. W. and others and {The Muon (g-2) Collaboration}},
  title     = {Search for Lorentz and CPT violation effects in muon spin precession},
  journal   = {Physical Review Letters},
  volume    = {100},
  issue     = {9},
  pages     = {091602},
  year      = {2008},
  doi       = {10.1103/PhysRevLett.100.091602}
}

@article{Battat2007,
  author    = {Battat, J. B. R. and Chandler, J. F. and Stubbs, C. W.},
  title     = {Testing for Lorentz Violation with Lunar Laser Ranging},
  journal   = {Physical Review Letters},
  volume    = {99},
  issue     = {24},
  pages     = {241103},
  year      = {2007},
  doi       = {10.1103/PhysRevLett.99.241103}
}

@article{Armitage2018,
  author    = {Armitage, N. P. and Mele, E. J. and Vishwanath, Ashvin},
  title     = {Weyl and Dirac semimetals in three-dimensional solids},
  journal   = {Reviews of Modern Physics},
  volume    = {90},
  issue     = {1},
  pages     = {015001},
  year      = {2018},
  month     = {Jan},
  doi       = {10.1103/RevModPhys.90.015001},
  publisher = {American Physical Society}
}

@article{Soluyanov2015,
  author    = {Soluyanov, Alexey A. and Gresch, Dominik and Wang, Zhijun and Wu, QuanSheng and Troyer, Matthias and Dai, Xi and Bernevig, B. Andrei},
  title     = {Type-II Weyl semimetals},
  journal   = {Nature},
  volume    = {527},
  number    = {7579},
  pages     = {495--498},
  year      = {2015},
  month     = {Nov},
  doi       = {10.1038/nature15768}
}

@article{Burkov2014,
  author    = {Burkov, A. A.},
  title     = {Anomalous Hall Effect in Weyl Metals},
  journal   = {Physical Review Letters},
  volume    = {113},
  issue     = {18},
  pages     = {187202},
  year      = {2014},
  month     = {Oct},
  doi       = {10.1103/PhysRevLett.113.187202},
  publisher = {American Physical Society}
}

@article{Xiong2015,
  author    = {Xiong, Jun and Kushwaha, Satya K. and Liang, Tian and Krizan, Jason W. and Hirschberger, Max and Wang, Wudi and Cava, R. J. and Ong, N. P.},
  title     = {Evidence for the chiral anomaly in the Dirac semimetal Na3Bi},
  journal   = {Science},
  volume    = {350},
  number    = {6259},
  pages     = {413--416},
  year      = {2015},
  month     = {Oct},
  doi       = {10.1126/science.aac6089}
}

@article{yang2018flexo,
  title={Flexo-photovoltaic effect},
  author={Yang, Ming-Min and Kim, Dong Jik and Alexe, Marin},
  journal={Science},
  volume={360},
  number={6391},
  pages={904--907},
  year={2018},
  publisher={American Association for the Advancement of Science}
}

@article{sipe2000second,
  title={Second-order optical response in semiconductors},
  author={Sipe, J. E. and Shkrebtii, A. I.},
  journal={Phys. Rev. B},
  volume={61},
  pages={5337},
  year={2000}
}

@article{BELICH2,
title = {A comment on the topological phase for anti-particles in a Lorentz-violating environment},
journal = {Physics Letters B},
volume = {639},
number = {6},
pages = {675-678},
year = {2006},
issn = {0370-2693},
doi = {https://doi.org/10.1016/j.physletb.2006.07.003},
url = {https://www.sciencedirect.com/science/article/pii/S0370269306008446},
author = {H. Belich and T. Costa-Soares and M.M. Ferreira and J.A. Helayël-Neto and M.T.D. Orlando},
}

@article{BELICH1,
title = {Classical solutions in a Lorentz violating scenario of Maxwell-Chern-Simons-Proca electrodynamics},
journal = {The European Physical Journal C},
volume = {42},
number = {1},
pages = {127-137},
year = {2005},
issn = {1434-6052},
author = {H. Belich and T. Costa-Soares and M.M. Ferreira and J.A. Helayël-Neto and M.T.D. Orlando},
}

@article{BELICH3,
  title = {Lorentz-violating corrections on the hydrogen spectrum induced by a nonminimal coupling},
  author = {Belich, H. and Costa-Soares, T. and Ferreira, M. M. and Helay\"el-Neto, J. A. and Moucherek, F. M. O.},
  journal = {Phys. Rev. D},
  volume = {74},
  issue = {6},
  pages = {065009},
  numpages = {6},
  year = {2006},
  month = {Sep},
  publisher = {American Physical Society},
  doi = {10.1103/PhysRevD.74.065009},
  url = {https://link.aps.org/doi/10.1103/PhysRevD.74.065009}
}

@article{jackiw,
  title = {Limits on a Lorentz- and parity-violating modification of electrodynamics},
  author = {Carroll, Sean M. and Field, George B. and Jackiw, Roman},
  journal = {Phys. Rev. D},
  volume = {41},
  issue = {4},
  pages = {1231--1240},
  numpages = {0},
  year = {1990},
  month = {Feb},
  publisher = {American Physical Society},
  doi = {10.1103/PhysRevD.41.1231},
  url = {https://link.aps.org/doi/10.1103/PhysRevD.41.1231}
}

@article{xue_ws2_2024,
	title = {{WS}$_{\textrm{2}}$ ribbon arrays with defined chirality and coherent polarity},
	volume = {384},
	issn = {0036-8075, 1095-9203},
	url = {https://www.science.org/doi/10.1126/science.adn9476},
	doi = {10.1126/science.adn9476},
	language = {en},
	number = {6700},
	urldate = {2025-02-18},
	journal = {Science},
	author = {Xue, Guodong and Zhou, Ziqi and Guo, Quanlin and Zuo, Yonggang and Wei, Wenya and Yang, Jiashu and Yin, Peng and Zhang, Shuai and Zhong, Ding and You, Yilong and Sui, Xin and Liu, Chang and Wu, Muhong and Hong, Hao and Wang, Zhu-Jun and Gao, Peng and Li, Qunyang and Zhang, Libo and Yu, Dapeng and Ding, Feng and Wei, Zhongming and Liu, Can and Liu, Kaihui},
	month = jun,
	year = {2024},
	pages = {1100--1104},
}

@article{chen_bulk_2025,
	title = {Bulk photovoltaic effect in two-dimensional ferroelectric  semiconductor $\alpha$-{In2Se3}},
	volume = {10},
	language = {en},
	number = {4},
	urldate = {2025-02-06},
	journal = {Nanoscale},
	author = {Chen, Yanting and Liang, Lijia and Zhang, Shuqin and Huang, Dianshuai and Zhang, Jing and Xu, Shuping and Liang, Chongyang and Xu, Weiqing},
	year = {2025},
	pages = {1622--1630},
}

@article{pal_bulk_2021,
	title = {Bulk photovoltaic effect in {BaTiO3}-based ferroelectric oxides: {An} experimental and theoretical study},
	volume = {129},
	issn = {0021-8979},
	shorttitle = {Bulk photovoltaic effect in {BaTiO3}-based ferroelectric oxides},
	url = {https://doi.org/10.1063/5.0036488},
	doi = {10.1063/5.0036488},
	number = {8},
	urldate = {2025-05-27},
	journal = {Journal of Applied Physics},
	author = {Pal, Subhajit and Muthukrishnan, S. and Sadhukhan, Banasree and N. V., Sarath and Murali, D. and Murugavel, Pattukkannu},
	month = feb,
	year = {2021},
	pages = {084106},
}

@article{li_enhanced_2021,
	title = {Enhanced bulk photovoltaic effect in two-dimensional ferroelectric {CuInP2S6}},
	volume = {12},
	copyright = {2021 The Author(s)},
	issn = {2041-1723},
	url = {https://www.nature.com/articles/s41467-021-26200-3},
	doi = {10.1038/s41467-021-26200-3},
	language = {en},
	number = {1},
	urldate = {2025-06-05},
	journal = {Nature Communications},
	author = {Li, Yue and Fu, Jun and Mao, Xiaoyu and Chen, Chen and Liu, Heng and Gong, Ming and Zeng, Hualing},
	month = oct,
	year = {2021},
	note = {Publisher: Nature Publishing Group},
	pages = {5896},
}

@article{krishna_understanding_2023,
	title = {Understanding the large shift photocurrent of \$\{{\textbackslash}mathrm\{{WS}\}\}\_\{2\}\$ nanotubes: {A} comparative analysis with monolayers},
	volume = {108},
	shorttitle = {Understanding the large shift photocurrent of \$\{{\textbackslash}mathrm\{{WS}\}\}\_\{2\}\$ nanotubes},
	url = {https://link.aps.org/doi/10.1103/PhysRevB.108.165418},
	doi = {10.1103/PhysRevB.108.165418},
	number = {16},
	urldate = {2026-01-16},
	journal = {Physical Review B},
	author = {Krishna, Jyoti and Garcia-Goiricelaya, Peio and de Juan, Fernando and Ibañez-Azpiroz, Julen},
	month = oct,
	year = {2023},
	note = {Publisher: American Physical Society},
	pages = {165418},
}

@article{zhang_enhanced_2019,
	title = {Enhanced intrinsic photovoltaic effect in tungsten disulfide nanotubes},
	volume = {570},
	copyright = {2019 The Author(s), under exclusive licence to Springer Nature Limited},
	issn = {1476-4687},
	url = {https://www.nature.com/articles/s41586-019-1303-3},
	doi = {10.1038/s41586-019-1303-3},
	language = {en},
	number = {7761},
	urldate = {2026-01-16},
	journal = {Nature},
	author = {Zhang, Y. J. and Ideue, T. and Onga, M. and Qin, F. and Suzuki, R. and Zak, A. and Tenne, R. and Smet, J. H. and Iwasa, Y.},
	month = jun,
	year = {2019},
	note = {Publisher: Nature Publishing Group},
	pages = {349--353},
}
